\documentclass[10pt,aps,prd,floatfix,notitlepage,showpacs,nofootinbib,superscriptaddress]{revtex4-1}

\usepackage{latexsym}
\usepackage{epsfig}
\usepackage{epstopdf}
\usepackage{graphicx}
\usepackage{amssymb}
\usepackage{amsmath}
\usepackage{dcolumn}
\usepackage{bm}
\usepackage{color}
\usepackage{comment}

\graphicspath{{pictures/}}

\begin{document}

\title{Teleparallel quintessence with a nonminimal coupling to a boundary term}

\author{Sebastian Bahamonde}
  \email{sebastian.beltran.14@ucl.ac.uk}
  \affiliation{Department of Mathematics, University College London, Gower Street, London, WC1E 6BT, UK}

\author{Matthew Wright}
\email{matthew.wright.13@ucl.ac.uk}
\affiliation{Department of Mathematics, University College London, Gower Street, London, WC1E 6BT, UK}

\date{\today}

\begin{abstract}
We propose a new model in the teleparallel framework  where we consider a scalar field nonminimally coupled to both the torsion $T$ and a boundary term given by the divergence of the torsion vector $B=\frac{2}{e}\partial_\mu (eT^\mu)$. This is inspired by the relation $R=-T+B$ between the Ricci scalar of general relativity and the torsion of teleparallel gravity. This theory in suitable limits incorporates both the nonminimal coupling of a scalar field to torsion, and the nonminimal coupling of a scalar field to the Ricci scalar. We analyse the cosmology of such models, and we perform a dynamical systems analysis on the case when we have only a pure coupling to the boundary term. It is found that the system generically evolves to a late time accelerating attractor solution without requiring any fine tuning of the parameters. A dynamical crossing of the phantom barrier is also shown to be possible.  
\end{abstract}

\maketitle

\section{Introduction}

The late time acceleration of the universe is one of the great problems of modern physics. Ever since this acceleration was suggested by Type-Ia supernovae surveys \cite{Riess:1998cb,Perlmutter:1998np}, this important result has been backed up by ever increasingly precise cosmological observations, including measurements of the cosmic microwave background (CMB) \cite{Komatsu:2010fb,Ade:2013zuv}, the Hubble constant \cite{Riess:2009pu}, baryon acoustic oscillations \cite{Lampeitl:2009jq}, and further measurements of Type-Ia supernovae \cite{Kowalski:2008ez}. Despite all these observations, there is a severe lack of understanding of this acceleration from a theoretical point of view. The standard cosmological model assumes the existence of a cosmological constant to explain this acceleration. However, a cosmological constant suffers from theoretical difficulties dues to its extremely small observed value compared to predictions from quantum field theoretical considerations, see~\cite{Martin:2012bt} for a review of this subject.

Teleparallel gravity is an alternative formulation of gravity which is equivalent to general relativity. General relativity is a geometric theory based on the Levi-Civita connection, which possesses curvature but zero torsion. On the other hand teleparallel gravity uses a different connection. There is a result originally discovered by Weitzenb\"ock that it is always possible to define a connection on a space such that is globally flat, in other words it possesses zero curvature. This connection is called the Weitzenb\"ock connection, and although it has a vanishing Ricci tensor it has non trivial torsion. This result is used to formulate an action based on a gravitational scalar called the torsion scalar $T$, which uses the Weitzenb\"ock connection. The dynamics of this action are completely equivalent to general relativity, and this follows from the following result
\begin{align}
R=-T+B
\end{align}
where $R$ is the Ricci Scalar, and $B$ is a boundary term related to the divergence of the torsion. Since $B$ is a total derivative, it gives no contribution to the field equations, and hence the action of teleparallel gravity is completely equivalent to the Einstein Hilbert action. 

There are numerous ways to modify both general relativity and teleparallel gravity with the aim of being able to understand the late time acceleration of the universe without the need for a cosmological constant. One approach is to modify the gravitational sector and one can consider both $f(R)$ theories of gravity and $f(T)$ theories. In general these give rise to different dynamics, since one can no longer write $f(R)=f(T) + \textrm{a total derivative}$. $f(R)$ gravity generically has fourth order field equations. In some sense $f(T)$ theories are less harmful in that this modification only results in second order field equations, however the price you pay for this is that local Lorentz invariance is lost.  In a recent paper~\cite{Us} a more general theory was considered in the teleparallel framework called $f(T,B)$ gravity, where $B$ is a divergence of the torsion, which incorporates both $f(R)$ and $f(T)$ gravity as particular subcases. 

Another approach is to modify the contents of the universe and add a form of matter into the universe called dark energy which possesses a negative pressure. This can be achieved by adding to the matter sector a canonical scalar field~\cite{Ratra:1987rm,Wetterich:1987fm,Zlatev:1998tr,Boisseau:2000pr,
Guo:2006ab,Dutta:2009yb,Sahni:1998at,Uzan:1999ch,Faraoni:2000wk,
Gong:2002dk,Elizalde:2004mq,Faraoni:2004dn,Setare:2008pc,Tamanini:2014mpa}, known as quintessence, a phantom scalar field~\cite{Caldwell2002a,Caldwell:2003vq,Nojiri:2003vn,Onemli:2004mb,Saridakis:2009pj,Dutta:2009dr}, or a combination of both of these fields called quintom~\cite{Feng:2004ad,Guo:2004fq,Feng:2004ff,Zhao:2006mp,Lazkoz:2006pa,
Lazkoz:2007mx,Saridakis:2009jq,Setare:2008si,Cai:2009zp,Chimento:2008ws}. A review of these models can be found in~\cite{Copeland:2006wr} and \cite{Leon:2009ce}. 

One can also consider a coupling between the scalar field and the gravitational sector. The standard approach is to consider a coupling between the scalar field and the Ricci scalar, of the form $\xi R \phi^2$~\cite{Chernikov:1968zm,Birrell:1979ip,Callan:1970ze}. Such a nonminimal coupling has motivations from different contexts. It appears as a result of quantum corrections to the scalar field in curved spacetimes~\cite{Birrell:1984,Ford:1986sy} and it is also required by renormalisation considerations~\cite{Callan:1970ze}. It also appears in the context of superstring theories~\cite{Maeda:1985bq}. Such models have attempted to explain the early time inflationary epoch, however the simple model with a quadratic scalar potential is now disfavoured by the current Planck data~\cite{Planck:2013jfk,Ade:2015lrj,Martin:2013nzq}. 

In recent years an alternative formulation has been considered where the coupling occurs between the scalar field and torsion of the form $\xi T \phi^2$~\cite{Geng:2011aj,Xu:2012jf,Geng:2013uga,Skugoreva:2014ena,Otalora:2013tba,Geng:2011ka,Geng:2012vn,Wei:2011yr,Kucukakca:2013mya,Kucukakca:2014vja,Jamil,Bamba}. This gives rise to different dynamics and interesting phenomenology, for example phantom behaviour and dynamical crossing of the phantom barrier. A dynamical systems analysis of these models were considered in~\cite{Xu:2012jf}, and the observational constraints on such models were found in~\cite{Geng:2011ka}. Other types of nonminimal coupling to the torsion sector have been considered, for example in~\cite{Kofinas:2015hla} a coupling between torsion and derivatives of the scalar field are considered. 

In this paper we consider a different approach where we examine a nonminimal coupling between the scalar field to both the torsion scalar $T$ and the boundary term $B$. We note that coupling a scalar field to a boundary term is not a new idea, for example a nonminimal coupling between a scalar field boundary terms such as the Gauss-Bonnet term and higher Lovelock polynomials have previously been considered~\cite{Nojiri:2006je,Granda:2014zea}. Our theory encompasses both nonminimally coupled teleparallel gravity and nonminimally coupled general relativity in suitable limits.

We then consider in detail the dynamics of the model when we have only a coupling between the boundary term $B$ and the scalar field. We use dynamical systems techniques to examine the global dynamics of the system. These techniques are widely used in cosmological applications since they allow one to study all possible evolutional paths when there is no possibility of finding an exact solution.  

This paper is organised as follows. In Section~\ref{tele} we introduce our conventions and the teleparallel equivalent of general relativity. In Section~\ref{nmcsf} we briefly review previous models involving nonminimally coupled scalar fields, and introduce the Lagrangian and field equations of our model. In Section~\ref{Cosmology} we derive the equations determining the background cosmology. Finally in Section~\ref{dsboundary} we perform a dynamical systems analysis in the case where we have purely a coupling between the scalar field and the boundary term.

\section{Teleparallel equivalent of general relativity}
\label{tele}

In this section we introduce the teleparallel equivalent of general relativity. The dynamical variable in this theory are given by the tetrad $e^{a}_{\mu}$, where Greek indices are spacetime indices, and latin indices are tangent space indices. The metric $g_{\mu\nu}$ is given in terms of the tetrad as
\begin{align}
g_{\mu\nu}=e^{a}_{\mu}e^{b}_{\nu}\eta_{ab},
\end{align} 
where $\eta_{ab}$ is the Minkowski metric. We also introduce the inverse tetrad $E^{\mu}_{a}$, such that
\begin{align}
E_{m}^{\mu}e_{\mu}^{n}=\delta^{n}_{m}, \quad {\text {and}} \quad
E_{m}^{\nu}e_{\mu}^{m}=\delta^{\nu}_{\mu}.\label{deltamunu}
\end{align}
The quantity $e$ is defined to be the determinant of the tetrad $e^a_\mu$, and is equivalent to the volume element of the metric, $e=\sqrt{-g}$.

In what follows we will obey the conventions outlined in~\cite{Maluf:2013gaa}. In general relativity the Levi-Civita connection is used, however in contrast, in teleparallel gravity the Weitzenb\"ock connection $W_{\mu}\,^{a}\,_{\nu}$ is chosen, and is given by the following derivatives of the tetrad
\begin{align}
W_{\mu}{}^{a}{}_{\nu}=\partial_{\mu}e^{a}{}_{\nu}.
\end{align}
This connection possesses torsion but the Ricci tensor of this connection vanishes, and hence possesses zero curvature. The torsion tensor is simply the antisymmetric part of the Weitzenb\"ock connection
\begin{align}
T^{a}\,_{\mu\nu}&=W_{\mu}{}^{a}{}_{\nu}-W_{\nu}{}^{a}{}_{\mu}=\partial_{\mu}e_{\nu}^{a}-\partial_{\nu}e_{\mu}^{a}.
\end{align}
The tensor $T_{\mu}$, which we call the torsion vector, is defined as the unique nontrivial contraction of the torsion tensor
\begin{align}
T_{\mu}=T^{\lambda}\,_{\lambda\mu}.
\end{align}

Now the field equations of teleparallel gravity follow from varying the following Lagrangian density with respect to the tetrad
\begin{align}
\mathcal{L}_T= \frac{e}{2\kappa^2}S^{abc}T_{abc}, \label{taction}
\end{align}
where the tensor $S$ is defined as the following
\begin{align}
S^{abc}=\frac{1}{4}(T^{abc}-T^{bac}-T^{cab})+\frac{1}{2}(\eta^{ac}T^b-\eta^{ab}T^c).
\end{align}
This combination of $S^{abc}$ and $T_{abc}$ is usually denoted by $T$, which we call the torsion scalar
\begin{align}
	T=S^{abc}T_{abc}.
\end{align}

Now in order to show that teleparallel gravity is equivalent to general relativity, we must express quantities using the Weitzenb\"ock connection in terms of quantities involving the torsion free Levi-Civita connection. One can write the Levi-Civita connection ${}^0 \Gamma$ in terms of the Weitzenb\"ock connection as so
\begin{align}
{}^0 \Gamma^{\mu}_{\lambda\rho}=W_{\lambda}{}^{\mu}{}_{\rho}-K_{\lambda}{}^{\mu}{}_{\rho}, \label{Levi}
\end{align}
where here $K$ is called the contortion tensor and it is defined as
\begin{align}
2K_{\mu}\,^{\lambda}\,_{\nu}&=T^{\lambda}\,_{\mu\nu}-T_{\nu\mu}\,^{\lambda}+T_{\mu}\,^{\lambda}\,_{\nu}.
\end{align}
This contortion tensor is antisymmetric in its last two indices. Now expressing the Ricci scalar of the Levi-Civita connection in terms of the Weitzenb\"ock connection using~(\ref{Levi}), the following relation can be found
\begin{align}
  R = - T + \frac{2}{e}\partial_\mu (e T^\mu) \,. \label{ricciT}
\end{align}
As the difference between the Ricci scalar and the torsion scalar is simply a total derivative, the action~(\ref{taction}) gives rise to the same dynamics as the Einstein Hilbert action. This shows that teleparallel gravity is indeed equivalent to general relativity. Defining the boundary quantity
\begin{align}
B=\frac{2}{e}\partial_\mu (e T^\mu)
\end{align}
one then has simply the relation $R=-T+B$. In~\cite{Us} this led us to consider the modification $f(T,B)$ gravity. In this work we will follow a different approach to include the term $B$ nontrivially in the dynamics, by coupling it to a scalar field. We also note that one can write $B$ in terms of a Levi-Civita covariant derivative simply as $B=2\nabla_\mu T^\mu$.

\section{Non-minimally coupled scalar fields}
\label{nmcsf}

The first approach to nonminimally coupling a scalar field to the gravitational sector is to consider a coupling to the Ricci tensor as follows 
\begin{align}
 S = \int 
  \left[ 
    \frac{R}{2\kappa^2}+\frac{1}{2}(\partial_\mu \phi \partial^\mu \phi +\xi R \phi^2)-V(\phi) + L_{\rm m}
  \right] \sqrt{-g}\, d^4x \, .\label{actionricci}
\end{align}
This approach was originally considered in the context of Brans-Dicke theories motivated by introducing a variable gravitational constant depending on a scalar field. In this notation the effective gravitational constant can be written as
\begin{align}
G_{\rm eff}= \frac{G}{1+\kappa^2 \xi \phi^2},
\end{align}
and one can redefine such a scalar field so that it coincides with the standard Brans-Dicke field.

Non-minimally coupled quintessence corresponds to taking $\xi=0$ in the above Lagrangian. For a review of quintessence type models, see~\cite{Copeland:2006wr}. Quintessence alone can give rise to many interesting features from late time accelerated expansion of the universe to inflation~\cite{Ratra:1987rm,Wetterich:1987fm,Zlatev:1998tr}. However simple models of scalar field inflation are becoming disfavoured by the latest Planck data. Another issue with a simple quintessence approach is that the effective equation of state must always satisfy $w_{\rm eff}>-1$ and require a very flat fine tuned potential in order to explain current cosmological observations. 

When $\xi\neq 0$, the nonminimal coupling can be transformed to a minimal coupling via a conformal transformation from the Jordan frame to the Einstein frame. Such a transformation reduces the system to a quintessence model with a coupling between the scalar field and dark matter.  For a constant exponential potential, the system can then be written as a two dimensional dynamical system. Physical quantities in this frame can then be transformed back into physical quantities in the Jordan frame. For a review of the dynamics of these models, see chapter 9 of~\cite{Copeland:2006wr} and references therein. Alternatively, one can work directly in the Jordan frame; a dynamical systems analysis for various potentials have been considered by various authors, see~\cite{Hrycyna:2015eta,Sami:2012uh,Hrycyna:2009zj,Szydlowski:2008in} and references within.

An alternative approach has been to consider a scalar field nonminimally coupled to torsion, giving rise to teleparallel dark energy theories~\cite{Geng:2011aj}. The following action is considered
\begin{align}
 S = \int 
  \left[ 
    -\frac{T}{2\kappa^2}+\frac{1}{2}(\partial_\mu \phi \partial^\mu \phi -\xi T \phi^2)-V(\phi) + L_{\rm m}
  \right] e\, d^4x \,.\label{actiontorsio}
\end{align}
This gives rise to different dynamics to the case of the nonminimal coupling to the Ricci scalar. Of course with a minimal coupling, setting $\xi=0$, the two theories again become equivalent due to the teleparallel equivalence. This theory again has a richer structure than simple standard quintessence behaviour, with both phantom and quintessence type dynamics possible, along with dynamical crossing of the phantom barrier.

The equivalence between general relativity and teleparallel gravity breaks down as soon as one nonminimally couples a scalar field, the field equations result in different dynamics. In this paper, we consider a more general action~\cite{Bamba}, with the aim of unifying both of the previous considered approaches
\begin{align}
 S = \int 
  \left[ 
    -\frac{T}{2\kappa^2}+\frac{1}{2}(\partial_\mu \phi \partial^\mu \phi -\xi T \phi^2-\chi B \phi^2)-V(\phi) + L_{\rm m}
  \right] e\, d^4x.\label{action}
\end{align}
When one sets $\chi=-\xi$ one will recover an action which is equivalent to~(\ref{actionricci}), and when one sets $\chi=0$ the action~(\ref{actiontorsio}) is recovered. A particularly interesting subcase will be when $\xi=0$, corresponding to a pure coupling between the boundary term. Such a coupling has not been previously studied in the literature. One could in principle choose a more general coupling $\eta(\phi) B$ between the potential and the boundary, however the choice $\eta(\phi)=\phi^2$ ensures that the constant $\chi$ is dimensionless.

We now derive the field equations of the action~(\ref{action}). Varying the action with respect to the tetrad field yields the following field equations
\begin{align}
-\left(\frac{2}{\kappa^2}+2\xi \phi^2  \right)\left[ e^{-1}\partial_\mu (e S_{a}{}^{\mu\nu})-E_{a}^{\lambda}T^{\rho}{}_{\mu\lambda}S_{\rho}{}^{\nu\mu}-\frac{1}{4}E^{\nu}_{a}T\right]-E^{\nu}_a \left[\frac{1}{2}\partial_\mu \phi \partial^\mu \phi -V(\phi)\right]
\nonumber\\ +E^{\mu}_a \partial^\nu \phi \partial_\mu \phi - 4(\xi+\chi) E^\rho_a S_{\rho}{}^{\mu\nu}\phi \partial_\mu \phi-\chi\Big[E^{\nu}_{a}\Box(\phi^2)-E^\mu_a \nabla^{\nu}\nabla_{\mu}(\phi^2)\Big]= T^\nu_a, \label{fieldeqn}
\end{align}
where $\Box=\nabla^{\mu}\nabla_{\mu}$ and $\nabla$ is the covariant derivative with respect to the Levi-Civita connection. Here $T^\nu_a$ is the standard energy momentum tensor derived from varying the matter sector, and is not to be confused with torsion. 

The term in the field equations proportional to $S_{\rho}{}^{\mu\nu}$ is Lorentz violating~\cite{Li:2010cg}. This term vanishes only in the case when we have $\xi=-\chi$, which corresponds to the case where the nonminimal coupling reduces to that of a coupling to only the Ricci scalar. This result is along similar lines to that obtained in~\cite{Us}, where the modification $f(T,B)$ was considered, where it was found that the field equations are invariant under local Lorentz transformations if and only if $f(T,B)=f(-T+B)=f(R)$.

Now it can be shown that the Einstein tensor of the Levi-Civita connection can be related to the torsion sector via the relation
\begin{align}
G^{\sigma}{}_{\nu}=-\Big[2e^{-1}\partial_\mu (e S_{a}{}^{\mu\nu})-2E_{a}^{\lambda}T^{\rho}{}_{\mu\lambda}S_{\rho}{}^{\nu\mu}-\frac{1}{2}E^{\nu}_{a}T\Big]e^a_\sigma.
\end{align}
This means we can write the field equations in a covariant form as follows
\begin{align}
\left(\frac{2}{\kappa^2}+2\xi \phi^2  \right)G_{\mu\nu}-g_{\mu\nu} \left[\frac{1}{2}\nabla_\lambda \phi \nabla^\lambda \phi -V(\phi)\right]+ \nabla_\mu \phi \nabla_\nu \phi&
\nonumber\\  - 4(\xi+\chi) S_{\mu}{}^{\lambda}\,_{\nu}\phi \partial_\lambda \phi-\chi\Big[g_{\mu\nu}\Box(\phi^2)- \nabla_{\mu}\nabla_{\nu}(\phi^2)\Big]&= T_{\mu\nu}. \label{fieldeqncov}
\end{align}
It is readily seen that this equation reduces to the correct field equation for a nonminimal coupling of a scalar field to the Ricci scalar when one takes $\chi=-\xi$.

Finally we have the modified scalar field equation. This is obtained by varying the action with respect to the scalar field, yielding the following Klein-Gordon equation
\begin{align}
\Box \phi+V'(\phi)=(\xi T+\chi B)\phi. \label{KG}
\end{align}

\section{Cosmology}
\label{Cosmology}
In this section we will derive the background equations for the cosmology of the above models. We will consider the standard  spatially flat Friedmann-Robertson-Walker (FRW) tetrad $e^a_\mu=\textrm{diag}(1,a(t),a(t),a(t))$ corresponding to a spatially flat FRW metric
\begin{align}
ds^2=dt^2-a(t)^2(dx^2+dy^2+dz^2),
\end{align}
where $a(t)$ is the scale factor. We will assume the energy momentum tensor of the matter sector is standard barotropic matter given by an isotropic perfect fluid
\begin{align}
T^{\mu}_{\nu}={\rm diag}(\rho,-p,-p,-p),
\end{align} 
where $\rho$ is the matter energy density and $p$ is the pressure. We will also assume all dynamical quantities, including the scalar field $\phi$, are homogeneous, depending only the time $t$.  

Inserting this FRW tetrad into the field equations~(\ref{fieldeqn}) gives us the following Friedmann equations
\begin{align}
3H^2&= \kappa^2 \left( \rho+  \rho_\phi \right), \label{FE1}
\\
3H^2+2\dot{H}&=-\kappa^2\left(p+p_\phi  \right). \label{FE2}
\end{align}
Here $H$ is the Hubble parameter $H=\frac{\dot{a}}{a}$ and we have defined the energy density and pressure of the scalar field as follows
\begin{align}
\rho_\phi&=\frac{1}{2}\dot{\phi}^2+V(\phi)-3\xi H^2 \phi^2+6\chi H\phi\dot{\phi},
\\ p_\phi&=\frac{1}{2}(1-4\chi)\dot{\phi}^2-V(\phi)+ 2H \phi\dot{\phi}(2\xi+3\chi)+3H^2\phi^2(\xi+8\chi^2)+2 \phi^2\dot{H}(\xi+6\chi^2) +2\chi\phi V'(\phi)+12 \chi  H^2 \phi^2 (\xi +\chi ).
\end{align}
And the Klein-Gordon equation~(\ref{KG}) reduces to
\begin{align}
\ddot{\phi}+3H\dot{\phi}+ 6(\xi H^2+\chi (3H^2+\dot{H}))\phi   +V'(\phi)=0.
\end{align}
In the above derivations we have used that the torsion scalar and boundary term can be written as
\begin{align}
T=-6H^2, \quad B=-18H^2-6\dot{H}.
\end{align}

We see that the coupling to the torsion scalar introduces a term proportional to $\phi^2$ in the first Friedmann equation~(\ref{FE1}), whereas the boundary term involves the addition of a $\phi \dot{\phi}$ term. One can define the equation of state of the dark energy or scalar field as the following ratio of the scalar field pressure and energy density
\begin{align}
w_\phi=\frac{p_\phi}{\rho_\phi}.
\end{align}
And also we define the total or effective equation of state as
\begin{align}
w_{\rm eff}=\frac{p+p_\phi}{\rho+\rho_\phi}.
\end{align}
Analogously to the standard matter energy density
\begin{align}
\Omega_m=\frac{\kappa^2 \rho}{3H^2},
\end{align}
we will define the density parameter of the dark energy or scalar field as
\begin{align}
\Omega_\phi=\frac{\kappa^2 \rho_{\phi}}{3H^2}.
\end{align}
so that the relation $1=\Omega_m+\Omega_\phi$ holds. 

In this model matter obeys the standard conservation equations
\begin{align}
\dot{\rho}+3H (\rho+p)=0,
\end{align}
and hence also
\begin{align}
\dot{\rho_\phi}+3H (\rho_\phi+p_\phi)=-0.
\end{align}

\section{Nonminimal coupling purely to the boundary term}
\label{dsboundary}
In this section we analyse in detail using dynamical systems techniques the case where $\xi=0$, where we purely have a coupling of the scalar field to the boundary term. In this case the quantity $Q_\phi$ is always positive, and so there is an energy flow from dark matter to dark energy. 

Let us introduce the dimensionless variables
\begin{align}
\sigma^2=\frac{\kappa^2\rho}{3H^2}\,, \quad x^2=\frac{\kappa^2\dot\phi^2}{6H^2}\,,\quad
y^2=\frac{\kappa^2V}{3H^2}\,,\quad
z= 2\sqrt{6}\kappa  \chi\phi \,.
\label{variables}
\end{align}
which straightforwardly generalise the normalised variables used to analyse standard quintessence~\cite{Copeland:1997et}. The first Friedman equation~(\ref{FE1}) written in these variables is simply the constraint
\begin{align}
1=\sigma^2 +x^2+y^2+x z, \label{phase}
\end{align}
which will define the boundary of our phase space.

The phase space will be three dimensional, and we choose to work with the variables $x$, $y$, and $z$. Since the energy density of matter is non-negative, the relation
\begin{align}
x^2+y^2+ xz \leq 1 \label{phasespace}
\end{align}
must be satisfied. As in standard quintessence, due to symmetries of the system, we can assume without loss of generality that our potential is positive and so we only need to consider $y>0$. There is no restriction on the sign of $x$ or $z$, since $\dot{\phi}$ can be positive or negative, as can $\chi$. This means that generically our phase space is non-compact, except in the case of a minimal coupling with $\chi=0$. If we further introduce the variables $x=u+v$ and $z=-2v$ then the phase space becomes
\begin{align}
u^2-v^2+y^2\leq 1,
\end{align}
which we see is simply hyperbolic space $\mathbb{H}^2$.

In what follows we will assume that  the pressure is linearly related to the energy density via the standard equation of state, $p=w\rho$, with $w$ the constant matter equation of state parameter. $w$ is physically constrained to lie between $w=0$ corresponding to (dark) matter and $w=1/3$ corresponding to radiation.

We define the quantity $N=\ln a$ and denote derivatives with respect to $N$ by a prime
\begin{align}
x'=\frac{dx}{dN}=\frac{1}{H}\frac{dx}{dt}.
\end{align}
Now in these variables, the equations of motion can be written as the following autonomous system of first order differential equations
\begin{align}
 x'&=-\frac{ \sqrt{6} y^2 \lambda  (x z-2)+3(2 x+z) \left( x^2 (4 \chi +w -1)+x (w -1) z+ y^2 (w +1)-w +1\right)}{ \left(z^2+4\right)}, \label{ds1}\\
  y' &=-\frac{y \left(\sqrt{6} \lambda \left(x \left(z^2+4\right)+2 y^2 z\right)+4 \left(3 x^2 (w -1)+3 x (w -1) z+3 y^2 (w +1)\right)-6 (2w +2+z^2)\right)}{2 \left(z^2+4\right)}, \label{ds2}\\
  z' &= 12 \chi x \label{ds3}.
\end{align}
Here we have defined the quantity $\lambda$ by
\begin{align}
\lambda=-\frac{V'(\phi)}{\kappa V(\phi)}.
\end{align}
In order to close the system we will have to specify a form of $\lambda$. For the autonomous system to remain three dimensional, one needs to choose a form of the potential such that $\lambda$ can be written in terms of the variables $x$, $y$ and $z$. In this work we will assume the potential $V$ to have an exponential form of the kind
\begin{align}
V(\phi) = V_0\, e^{-\lambda\kappa\phi} \,,
\label{055}
\end{align}
with $V_0$ a constant. This ensures that $\lambda$ is simply a constant. However this is not the only choice that will give a closed three dimensional system. One could also consider a power law potential of the form
\begin{align}
V(\phi)=\frac{M^{\alpha+4}}{\phi^\alpha},
\end{align}
where $\alpha$ is a constant and $M$ is a positive constant with the units of mass. In the context of late time acceleration of the universe, $\alpha$ is usually taken to be positive, however when considering inflation negative $\alpha$ is often considered. In either case, this would allow one to write $\lambda$ in terms of $z$ as
\begin{align}
\lambda=\frac{2\sqrt{6}\alpha\chi}{z},
\end{align}
however we will leave the analysis of such a potential for future work.

\subsection{Finite critical points}

Let us denote the above autonomous system~(\ref{ds1})-(\ref{ds3}) as
\begin{align}
x_i=f_i(x,y,z), \quad x_i=(x,y,z). \label{dynam}
\end{align}
Critical or fixed points of~(\ref{dynam}) correspond to $(x_*,y_*,z_*)$ that are solutions to all three equations $f_i(x_*,y_*,z_*)=0$. We note that the right hand side of the dynamical system~(\ref{ds1})-(\ref{ds3}) suffers from no singularities since the denominator is always well defined. 

We can rewrite physical quantities in terms of the new variables $x$, $y$ and $z$. We find for the effective equation of state that
\begin{align}
w_{\rm eff}=-\frac{12 x^2 (4 \chi +w -1)+12 (w -1) x z+2 y^2 \left(\sqrt{6} \lambda  z+6 w +6\right)-3 \left(z^2+4 w \right)}{3 \left(z^2+4\right)}.
\end{align}
Now at a fixed point, one can integrate the second Friedmann equation~(\ref{FE2}) explicitly to find $a$, and it is found
\begin{align}
a \propto (t-t_0)^{2/[3(1+w_{\rm eff})]} .\label{asol}
\end{align}
This means that the universe expansion will be accelerating if $w_{\rm eff}<-1/3$. We can also express the energy density of the matter and scalar field in terms of $x$ and $y$
\begin{align}
\Omega_m&=1-x^2-y^2-xz,
\\
\Omega_\phi &=x^2+y^2+xz.
\end{align}

\begin{table}[!ht]
\begin{tabular}{|c|c|c|c|c|}
  \hline
  Point & $x$ & $y$ & $z$ & Existence\\
  \hline \hline 
  O & 0 & 0 & 0 & $\forall \lambda, \, \chi$ \\
  \hline
  $A_{\pm}$ & $\pm 1$ & 0 & 0 & $\chi=0$ \\
  \hline
  $B$ & $\sqrt{\frac{3}{2}} \frac{(1+w)}{\lambda}$ & 
  $\sqrt{\frac{3}{2}} \frac{\sqrt{(1+w)(1-w)}}{\lambda}$ & 0 & $\chi=0$ and $\lambda ^2>3 (1+w)$ \\
  \hline
  $C$ & $\frac{\lambda}{\sqrt{6}}$ & $\sqrt{1-\frac{\lambda^2}{6}}$ &
  $0$ & $\chi=0$ and $\lambda^2<6$  \\
  \hline
  $D$ & $0$ & $1$ & $\sqrt{\frac{2}{3}} \lambda $ & $\forall \lambda, \, \chi$
  \\ \hline
  $E$ & $0$ &
  1 & $0$ &$\lambda=0$ and $\chi\neq 0$  \\
  \hline
\end{tabular}
\caption{Critical points of the autonomous system~(\ref{ds1})-(\ref{ds3}), along with the conditions for existence of the point. Points $A_{\pm}$, $B$ and $C$ exist only when $\chi=0$, corresponding to standard quintessence.}
\label{tab:m1_gen}
\end{table}

A list of the critical points, along with their condition for existence is displayed in Tab.~\ref{tab:m1_gen}. We emphasise that the points $A_{\pm}$, $B$ and $C$ appear only in the case when $\chi=0$, and are simply the standard critical points of quintessence models with an exponential potential. However, they are included in the table for completeness. In the generic case when $\chi \neq 0$ and $\lambda \neq 0$ the system has only two critical points given by $O$ and $D$. This is because the third equation of our dynamical system~(\ref{ds3}) forces $x=0$, and thus this severely restricts the possible solution set of the remaining two equations. In the case of a nonminimal coupling to the torsion $T$ there are in general three finite critical points, in this case the number is reduced to only two.

In order to determine the linear stability of these points, one must analyse the following Jacobian matrix of partial derivatives
\begin{align}
J=\frac{\partial f_i(x,y,z)}{\partial x_j}, \quad i,j=1,2,3
\end{align}
evaluated at each of the critical points. Information about the stability of each of the points is then contained in the eigenvalues of this matrix. The point will be stable if all three eigenvalues have negative real part, unstable if all three eigenvalues have positive real part, and will be a saddle point if there are both positive and negative eigenvalues. If one of the eigenvalues is zero then the critical point is said to be non-hyperbolic, and then one must go beyond a simple linear stability analysis treatment, for a discussion of how to do this, see for instance~\cite{Boehmer:2014vea}. However, all of our critical points are hyperbolic so such a treatment will not be needed here.

\begin{table}[!ht]
\begin{tabular}{|c|c|c|c|c|}
\hline
**\mbox{Point} &  $w_{\rm eff}$ & Acceleration & Eigenvalues & Stability \\
\hline \hline
$O$ & $0$ & No & $\frac{3}{2},-\frac{3}{4} \left(\sqrt{1-16 \chi }+1\right),\frac{3}{4} \left(\sqrt{1-16 \chi }-1\right)$& Saddle node\\
\hline
$A_{-}$ & $1$   & No & $3, \, 3+\sqrt{\frac{3}{2}} \lambda $ & Unstable node:  $\lambda>-\sqrt{6}$ \\
& &  & & Saddle node: otherwise \\
\hline
$A_{+}$ &  $1$ & No &$3, \,  3-\sqrt{\frac{3}{2}} \lambda$& Unstable node: $\lambda<\sqrt{6}$ \\
& & && Saddle node: otherwise \\
\hline
$B$ &  $0$ & No &$\frac{3}{4}+\frac{3 \sqrt{24-7 \lambda ^2}}{4 \lambda },-\frac{3}{4}+\frac{3  \sqrt{24-7 \lambda ^2}}{4 \lambda }$ & Stable node: $3<\lambda^2<24/7$  \\
& & & & Stable spiral: $\lambda^2>24/7$ 
\\ \hline $C$ & $\frac{\lambda^2-3}{3}$ & $\lambda^2<2$ & $\lambda ^2-3, \, \frac{1}{2} \left(\lambda ^2-6\right)$& Stable node: $\lambda^2<3  \:\&\:\lambda\beta>(\lambda^2-3)$\\ & & & & Saddle node: $\beta<(\lambda^2-3)/\lambda$ \\
\hline
$D$   & $-1$ & Yes & $-3, \frac{3}{2} \left(-1+\frac{\sqrt{\lambda ^2-48 \chi +6}}{\sqrt{\lambda ^2+6}}\right), $ & Stable spiral: $48\chi>\lambda^2+6$   \\
& & & $\frac{3}{2}\left(-1-\frac{\sqrt{\lambda ^2-48 \chi +6}}{\sqrt{\lambda ^2+6}}\right)$& Saddle point: $0<48\chi<\lambda^2+6$  \\
& & & & Stable point: $\chi<0$ \\
\hline
$E$  & $-1$ & Yes& $-3, \, -\frac{3}{2} \left(\sqrt{1-8 \chi }+1\right)$ &  Stable spiral: $\chi > 1/8$\\
& & &$\frac{3}{2} \left(\sqrt{1-8 \chi }-1\right)$ &  Stable node: $0<\chi<1/8$\\
& & & &  Saddle point: $\chi<0$
\\
\hline
\end{tabular}
\caption{Stability and eigenvalues of the critical points of the dynamical system~(\ref{ds1})-(\ref{ds3}) along with the effective equation of state assuming a (dark) matter equation of state $w=0$.} 
\label{tab01}
\end{table}

The eigenvalues and stability properties of the critical points are displayed in Tab.~\ref{tab01}. The points $A_{\pm}$, $B$ and $C$ exist only for $\chi=0$, in which case $z=0$ identically and the system reduces to a two dimensional system. Hence only the eigenvalues of the reduced two dimensional system are displayed. The eigenvalues of point $D$ mean that there are three three separate cases: for $0<48\chi<\lambda^2+6$ all three eigenvalues are negative and so the point is a stable node, if $48\chi>\lambda^2+6$ there is one negative and two imaginary eigenvalues (with negative real part) and so the point is a stable spiral, whereas for $\chi<0$ there is always at least one negative and one positive eigenvalue and hence the point is a saddle.   

\subsection{Critical points at infinity}
Now since the phase space is not compact, we must carefully check whether there are critical points at infinity.  An analysis of the dynamics at infinity is important to understand the global stability of the system. There are a number of approaches one can take to doing this. One approach is to use projective coordinates, a technique recently used in~\cite{Hrycyna:2015eta}. One could also use a function such as $\arctan$ to compactify the variables, see for example~\cite{Boehmer:2015kta}.

In this work we follow the approach used in~\cite{Xu:2012jf}, and use Poincar\'{e} variables to compactify the phase space~\cite{Lynch}. We begin by introducing the compactified coordinates $x_r$, $y_r$ and $z_r$ like so
\begin{align}
x_r= \frac{x}{\sqrt{1+r^2}}, \quad  y_r= \frac{y}{\sqrt{1+r^2}}, \quad z_r= \frac{z}{\sqrt{1+r^2}},
\end{align}
where $r^2=x^2+y^2+z^2$. We also define the quantity $\rho=\frac{r}{\sqrt{1+r^2}}$, so that $x_r^2+y_r^2+z_r^2=\rho^2$. This means the dynamics at infinity will now be captured by taking the limit $\rho\rightarrow 1$. We then make a further coordinate transformation, transforming the Poincar\'{e} variables into spherical polar coordinates as so
\begin{align}
x_r= \rho \cos \theta \sin \varphi, \quad z_r=\rho \sin \theta \sin \varphi, \quad y_r= \rho \cos \varphi,
\end{align}
where the variables lie in the range $\rho \in [0,1]$, $\theta \in [0,2\pi]$ and since we are restricting ourselves to $y\geq0$ the angle $\varphi$ lies in the restricted range $\varphi\in [-\frac{\pi}{2},\frac{\pi}{2}]$.

The Friedmann constraint~(\ref{phasespace}) can be written in Poincar\'{e} variables as
\begin{align}
2x_r^2+2y_r^2+ z_r^2 +x_r z_r \leq 1, \label{poincphase}
\end{align}
and hence the phase space will be the intersection of this ellipsoid with the Poincar\'{e} sphere
\begin{align}
x_r^2+y_r^2+z_r^2 \leq 1.\label{poinchpase2}
\end{align}

Transforming our dynamical system~(\ref{ds1})-(\ref{ds3}) into these new variables, in the limit $\rho\rightarrow 1$ we find the following
\begin{align}
\rho '&=0,
\\
\sqrt{1-\rho^2}\theta'&= \sqrt{6} \lambda  \cos \theta \cos \varphi  \cot \varphi,
\\
\sqrt{1-\rho^2}\varphi'&= \sqrt{\frac{3}{2}} \lambda \cos \varphi \left(2 \sin \theta \cos ^2\varphi+\cos \theta  \sin ^2\varphi\right),
\end{align}
and hence the angular part of the equation decouples. Setting the right hand side of these equations equal to zero, we find that we must have $\cos \varphi=0$, and hence the critical points are those at infinity which obey
\begin{align}
x_r&=\pm \cos\theta,\nonumber \\
z_r&=\pm \sin\theta, \nonumber \\
y_r&=0. \label{theta}
\end{align}

Now we can use this to find an equation for $\theta'$. We go back to the equations derived for $x_r', y_r'$ and $z_r'$ and insert the ansatz~(\ref{theta}), and then use the chain rule to find an expression for  $\theta'$
\begin{align}
\theta'=24 \chi  \cos \theta \cot\theta  (\sin \theta +\cos \theta )-\frac{3}{2} (2 \cot \theta +1) (\sin (2 \theta )+\cos (2 \theta )+1)\,. \label{thetaeqn}
\end{align}
Setting the right hand side of this equal to zero gives the critical points at infinity. 

%There are in general six solutions to this equation, however the found solutions are too long to express analytically. Moreover not all of these points at infinity obey the Friedmann constraint~(\ref{poincphase}), however determining how many of the solutions do is impossible analytically for a general $\chi$. 

In the Poincar\'{e} variables we have that the dark energy density parameter is given by
\begin{align}
\Omega_\phi &=\frac{x_{r}^2+y_{r}^2+x_{r}z_{r}}{1-x_{r}^2-y_{r}^2-z_{r}^2}.
\end{align}
At the critical points of~(\ref{thetaeqn}), this dark energy density parameter is divergent. Similarly, since the relation $1=\Omega_m+\Omega_\phi$ the matter energy density will also be divergent at this point, and so these points are not of physical interest~\cite{Xu:2012jf}, they are only of mathematical interest. Thus the model has no physically relevant critical points at infinity. Nonetheless, since our phase space is non-compact, and the boundary of our phase space is an abnormal shape, we will continue using the Poincar\'{e} variables when plotting the trajectories.

\subsection{Cosmological implications}

\begin{figure}[!ht]
\includegraphics[width=0.6\textwidth]{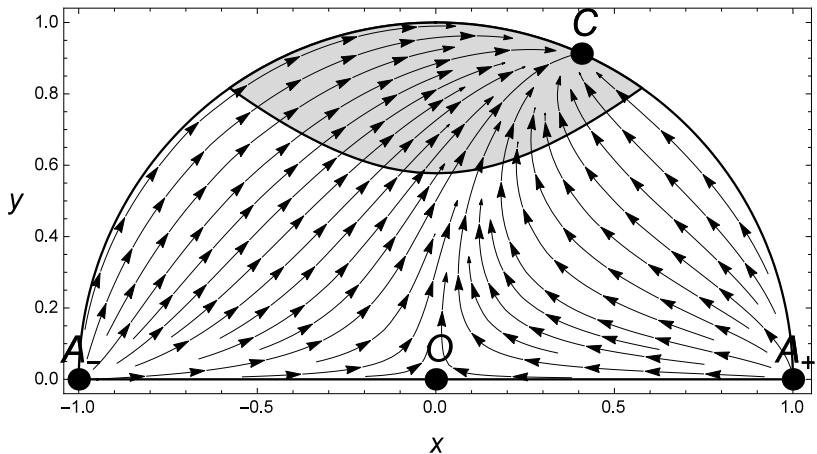}
\caption{Phase space showing trajectories of standard quintessence models, for the particular parameter choice $w=0$, $\lambda=1$ (with $\chi=0$). The point $C$ is the late time accelerating attractor, with the shaded region indicating the region of acceleration. }\label{quintessencepic}
\end{figure}

In this section we will discuss the dynamics of the above system. We will first briefly review the dynamics of minimally coupled quintessence and nonminimally coupled teleparallel energy in order to compare our results.

The minimally coupled quintessence is achieved by setting $\chi=0$ in the above model. The dynamical system is then two dimensional and the phase space is simply the upper half unit disc. This scenario has potentially five critical points, $O$, $A_{\pm}$, $B$ and $C$. Point $O$ is the matter dominated point, and it is always a saddle point, as required by cosmological observation. The points $A_{\pm}$ are both unphysical stiff matter points with effective equation of state $w_{\rm eff}=+1$. These points can either be saddle points or early time attractors depending on the value of the parameter $\lambda$. The point $B$ is of particular interest cosmologically since it is a scaling solution, the effective equation of state exactly mirrors the matter equation of state $w_{\rm eff}=w$, despite the fact the energy density of the scalar field does not vanish at this point. 

The point $C$ is entirely scalar field dominated, and exists only for $\lambda^2<6$. For a suitably flat potential $\lambda^2<2$, this point can describe an accelerating universe, however the effective equation of state is bounded below by $-1$, and thus crossing into the phantom regime is not possible, and $\lambda$ has to approach zero for $w_{\rm eff}$ to approach $-1$. For $\lambda^2<3$ the late time attractor is given by this point $C$, whereas for $\lambda^2>3$ point $B$ is the global attractor. Hence it is possible in these models to achieve a late time accelerating attracting solution. A typical plot of the two dimensional phase space of these quintessence models is displayed in Fig.~\ref{quintessencepic}, where the parameter choice $\lambda=1$ is chosen. Here we see that many trajectories pass close to the matter dominated origin before passing through the shaded acceleration region before ending at the late time attractor at point $C$. 

Teleparallel dark energy, on the other hand, means we must take $\chi=0$ but restore $\xi$ into our action~(\ref{action}). The phase space analysis was first explored in~\cite{Wei:2011yr}, and a further analysis, where critical points at infinity were taken into account was performed in~\cite{Xu:2012jf}, and it is this analysis we review here. For generic $\xi$ and $\lambda$, the system has three finite critical points, the origin $O$, along with two further points $F$ and $G$ (following the same point labels as in~\cite{Xu:2012jf}). Along with these finite points, there are a further four critical points at infinity, denoted by $K_{\pm}$ and $L_{\pm}$. 

The critical points at infinity are either saddle points or unstable, however the points $K_{\pm}$ can give rise to an accelerating universe, and thus these models are able to undergo transient inflationary periods. The points $F$ and $G$ both describe dark energy dominated points, with effective equation of state $w_{\rm eff}=-1$. The point $G$ is only a saddle node, however the point $D$ is stable for $\lambda^2<\xi$, and thus for a large class of parameters these models have a late time accelerating attractor, without requiring any fine tuning. One can also get $w_{\rm eff}=-1$ while still having an arbitrary value of the potential, which is a significant advantage over standard quintessence, which requires a very small $\lambda$ to achieve such an acceleration.

Now let us explore the features of our model. The points $A_{\pm}$, $B$ and $C$ only exist in the limit $\chi\rightarrow 0$, and of course the points exhibit the same behaviour as the above discussion. The point $O$ exists also for $\chi \neq 0$, and corresponds to a matter dominated universe with no scalar field contribution. This point remains a saddle point for all $\chi$ and $\lambda$. The model has a further two critical points $D$ and $E$. The point $E$ exists only when the parameter $\lambda=0$, meaning that the potential is simply a constant. This point is entirely dominated by the potential term, and simply corresponds to standard de Sitter type expansion, with the potential behaving exactly as a cosmological constant. This point also exists in the nonminimal teleparallel case and standard quintessence.

\begin{figure}[!ht]
\includegraphics[width=0.6\textwidth]{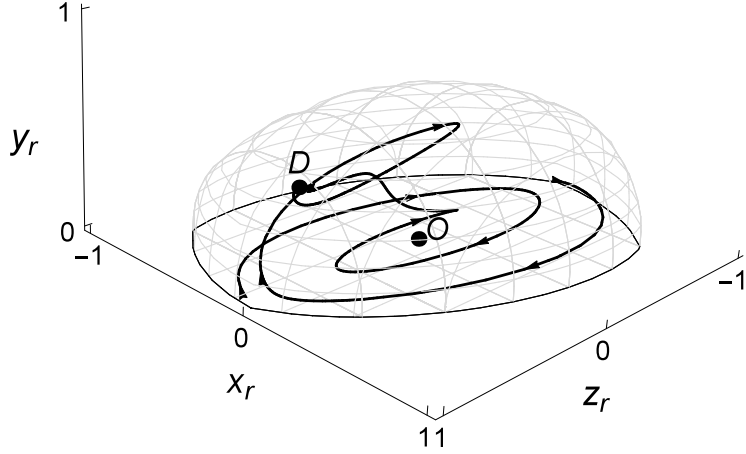}
\caption{Phase space showing trajectories of the dynamical system~(\ref{ds1})-(\ref{ds3}) in Poincar\'{e} variables when $\chi=1,\lambda=2$ and $w=0$. Point $D$ is the global attractor.}\label{chi1lambda2}
\end{figure}

The point $D$ exists only for $\chi \neq0$,  and so it is unique to this model, although its coordinates are independent of $\chi$.  At this point the energy density from the kinetic energy of the scalar field vanishes, but it has both contributions from both the matter sector and the potential energy of the scalar field. It has an effective equation of state $w_{\rm eff}=-1$ independent of the values of $\chi$ and $\lambda$. For positive $\chi$ this point is always a stable spiral, independent of $\lambda$ and hence it will always describe a late time accelerating attractor solution without requiring any fine tuning. 

In Fig.~\ref{chi1lambda2} we display some typical trajectories in the three dimensional Poincar\'{e} phase space for the particular parameter choice $\lambda=2$ and $\chi=1$. The boundary of the phase space is given in Poincar\'{e} coordinates by the intersection of~(\ref{poincphase}) and (\ref{poinchpase2}). Trajectories can pass close to the matter dominate origin $O$, before all trajectories end at the late time accelerating point $D$.

\begin{figure}[!ht]
\includegraphics[width=0.6\textwidth]{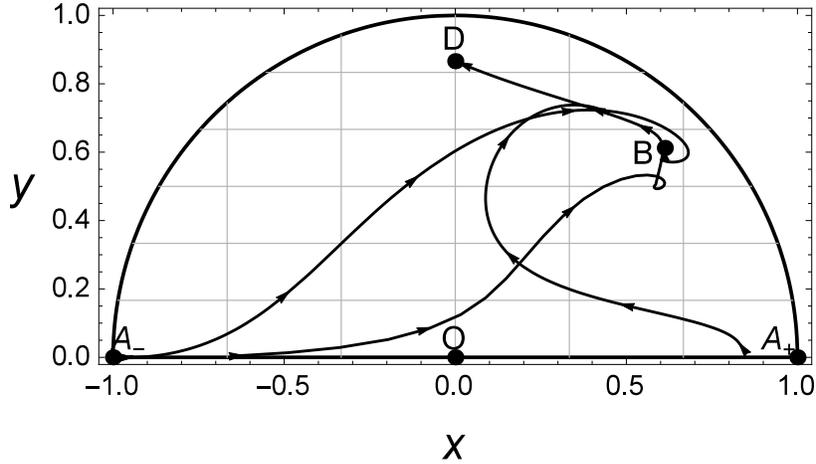}
\caption{Phase space showing trajectories of the dynamical system~(\ref{ds1})-(\ref{ds3}) projected onto the $x-y$ plane when $\chi=10^{-3},\lambda=2$ and $w=0$. The points $A_{\pm}$ and $B$ are quasi-stationary. Point $D$ is again the global attractor.}\label{lambda2ch10-3}
\end{figure}

In Fig.~\ref{lambda2ch10-3} we display a two dimensional projection onto the $x-y$ plane for the phase space when the parameter values are $\lambda=2$ and $\chi$ is chosen so that it is close to zero, $\chi=10^{-3}$. In this case the critical points of standard quintessence, points $A_{\pm}$, $B$ and $C$ behave as quasi-stationary points. Trajectories are still attracted close to these points. In the plot shown, trajectories start near the early time unstable points $A_{\pm}$ and are drawn towards the quasi scaling solution $B$. However, $B$ is no longer a true critical point and so trajectories then travel to the stable global attractor $D$, which is not present in standard quintessence. 

\begin{figure}[!ht]
\includegraphics[width=0.6\textwidth]{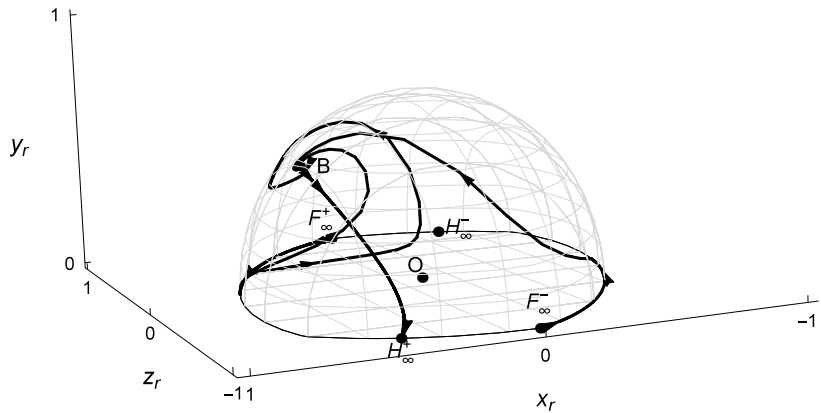}
\caption{Phase space showing trajectories of the dynamical system~(\ref{ds1})-(\ref{ds3}) when $\chi=-10^{-3},\lambda=2$ and $w=0$. Trajectories end at unphysical critical points lying at infinity.}\label{lambda2ch-1}
\end{figure}

The dynamics of the dynamical system are less interesting from a cosmological point of view when one considers a negative coupling $\chi$. In this case the point $D$ is no longer a global attractor, and trajectories are instead drawn towards the unphysical critical points at infinity. Such a scenario is displayed in Fig~\ref{lambda2ch-1}. The points $H_{\infty}^{\pm}$  lie on the boundary of the Poincar\'{e} sphere, and these are the mathematical critical points at infinity corresponding to the fixed point of equation~(\ref{thetaeqn}). Trajectories move towards the quasi-critical point $B$ before ending at one of these points at infinity, but as already noted above they are unphysical since both $\Omega_m$ and $\Omega_\phi$ are divergent at these points. Such models are therefore not physically viable.

\section{Discussion}

In this work we proposed introducing a nonminimal coupling of a scalar field to both the torsion scalar $T$ and the boundary term $B$. This model incorporates both nonminimal coupling to the Ricci scalar and a nonminimal coupling to the torsion scalar in suitable limits. We analysed in detail the dynamics of the background cosmology when we have simply a coupling to the boundary term. It is found that for a positive coupling, the system generically evolves to a late time dark energy dominated attractor, whose effective equation of state is exactly $-1$. This is independent of the potential, and thus requires absolutely no tuning of the potential to achieve this.

\begin{figure}[!ht]
\includegraphics[width=0.48\textwidth]{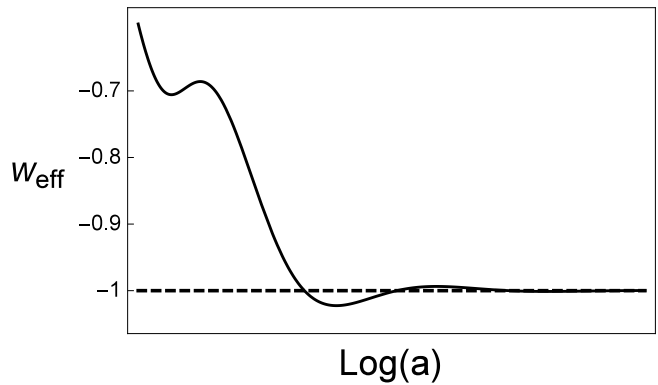}
\caption{Plot of $w_{\rm eff}$ against $N$ for a typical trajectory when the parameters $\lambda=2$, $\chi=1$ are chosen. The dashed line indicates the phantom barrier. }\label{wefflambda2chi1}
\end{figure}

Moreover while the system is evolving close to this late time attractor, the phantom barrier can indeed be crossed, a scenario impossible without the presence of the coupling. We display a plot typical of such behaviour in Fig.~\ref{wefflambda2chi1}. It is seen that the effective equation of state can cross the phantom barrier, and indeed cross from both directions, oscillating around the barrier before settling at its final late time de Sitter type expansion.

The global dynamics of these models are simpler than the case of the nonminimally coupled torsion scenario. There are less critical points, and there are no longer any physical critical points at infinity. Teleparallel dark energy also possesses saddle points describing an accelerating universe and hence can exhibit transient periods of inflation. Such a scenario is not possible in our model as we have no accelerating saddle points.

An investigation of the full phase space when a coupling to both $T$ and $B$ is present could in principle be analysed in future. In our case however we focused only on a coupling to the boundary term here to analyse the effects of this new contribution. One should also investigate the role solar system and other observational constraints place on these nonminimal coupled models, analogous to~\cite{Geng:2011ka}. In this work we focused on exploring the background cosmology and so furthermore the cosmological perturbations should also be investigated, along the lines of~\cite{Geng:2012vn}.

\begin{acknowledgments}
	SB is supported by the Comisi{\'o}n Nacional de Investigaci{\'o}n Cient{\'{\i}}fica y Tecnol{\'o}gica (Becas Chile Grant No.~72150066). We also wish to thank Christian Boehmer for insightful discussions and comments on the manuscript. 
\end{acknowledgments}

\end{document}